\documentclass[twocolumn,twoside]{article}
\usepackage{graphics}
\newcommand{\h}{\linebreak \hspace*{3ex}}
\newcommand{\hb}{\\ \hspace*{2ex}}

\begin{document}
\title{A TECHNIQUE FOR SIMULTANEOUS MEASUREMENT OF CIRCULAR AND LINEAR POLARIZATION WITH A SINGLE-CHANNEL POLARIMETER}
\author{S.V.\,Kolesnikov$^{1,2}$, V.V.\,Breus$^3$, N.N.\,Kiselev$^{1,4}$, I.L.\,Andronov$^3$\\[2mm]
\begin{tabular}{l}
$^1$ Crimean Astrophysical Observatory (CrAO), Krimea, Nauchny\\
$^2$ Astronomical Observatory, I.I. Mechnikov Odessa National University (ONU)\hb
T.G.Shevchenko Park, Odessa 65014 Ukraine\\
$^3$ Department of Mathematics, Physics and Astronomy, Odessa National\hb
Maritime University (ONMU), Mechnikova 34, Odessa 65029 Ukraine\\
$^4$ Main Astronomical Observatory, National Academy of Sciences of Ukraine\hb
(MAO NASU), Zabolotnogo 27, Kyiv, Ukraine \\[2mm]
\end{tabular}
}
\date{}
\maketitle
ABSTRACT.
We present a technique for simultaneous measurement of circular and linear polarization with the single-channel polarimeter, that is used in Crimean astrophysical observatory for many years. Methods and a computer program for data reduction is described. The algorithm is described,
which have been used for photo-polarimetric monitoring of various astronomical objects  cataclysmic variables, asteroids, comets.\\[0.5mm]

{\bf 1. Introduction}\\[0.5mm]

The development of devices capable of measuring circular polarization of light became possible in the middle of the 20$^{th}$ century after the appearance of achromatic retarders. For us it became feasible in the middle of the 80's, following the development of multicomponent symmetric achromatic retarders by V.A. Kucherov (1986). Shakhovskoy N.M. et al. (2001) described a technique for circular polarization measuring using the CrAO 2.6-m Shajn mirror telescope (SMT) with a single-channel photopolarimeter, which uses the high-speed rotation of a quarter-wave retarder as a modulator. In that case the signal was integrated by four pulse counters over the time intervals corresponding to the angles of the retarder rotation $90^\circ$; at that, the "angles of activity" of the second pair of the pulse counters were shifted relative to those of the first pair by 45 degrees. In 2002, we modified the SMT polarimeter aiming to quasi-simultaneously measure all four Stokes parameters, namely I, Q, U and V. We still use a quarter-wave retarder as an analyser, which is continuously rotating at the rate of 33 rps, and a Glan prism as a stationary polarizer. The switching unit of the polarimeter has been rearranged in such a way that the signal is integrated over the time intervals corresponding to the 22.5 degree angles of the retarder rotation, i.e. by eight pulse counters.

According to Serkowski (1974), the intensity of the light beam which has passed through such a device can be expressed by the following formula:
\begin{equation}
I'(\psi)=\frac{1}{2}(I+\frac{Q}{2}(1+\cos4 \psi)+\frac{U}{2}\sin 4\psi-V\sin 2\psi)
\label{form1}
\end{equation}
where $\psi$ is the angle of rotation of the retarder fast axis (hereinafter the major axis) relative to the analyser principal plane; $I$ is the intensity of the incoming radiation; $Q$ and $U$ are the linear polarization parameters; and $V$ is the circular polarization parameter.
Having the dependencies (\ref{form1}) angularly integrated the equations for eight pulse counters ($0^{\circ}-22.5^{\circ}$, $22.5^{\circ}-45^{\circ}$, ..., $337.5^{\circ}-360^{\circ}$) and taking into account repeating of counts in the range ($180^{\circ}-360^{\circ}$), the following expected values may be derived:
\begin{eqnarray}
n_1=I_0\cdot\frac{\pi}{8}+\frac{Q}{2}\cdot\frac{\pi}{8}+\frac{Q}{2}\cdot\frac{1}{4}+\frac{U}{2}\cdot\frac{1}{4}-V\cdot(\frac{1}{2}-\frac{\sqrt{2}}{4})\nonumber\\
n_2=I_0\cdot\frac{\pi}{8}+\frac{Q}{2}\cdot\frac{\pi}{8}-\frac{Q}{2}\cdot\frac{1}{4}+\frac{U}{2}\cdot\frac{1}{4}-V\cdot\frac{\sqrt{2}}{4}\nonumber\\
n_3=I_0\cdot\frac{\pi}{8}+\frac{Q}{2}\cdot\frac{\pi}{8}-\frac{Q}{2}\cdot\frac{1}{4}-\frac{U}{2}\cdot\frac{1}{4}-V\cdot\frac{\sqrt{2}}{4}\nonumber\\
n_4=I_0\cdot\frac{\pi}{8}+\frac{Q}{2}\cdot\frac{\pi}{8}+\frac{Q}{2}\cdot\frac{1}{4}-\frac{U}{2}\cdot\frac{1}{4}-V\cdot(\frac{1}{2}-\frac{\sqrt{2}}{4})\\
n_5=I_0\cdot\frac{\pi}{8}+\frac{Q}{2}\cdot\frac{\pi}{8}+\frac{Q}{2}\cdot\frac{1}{4}+\frac{U}{2}\cdot\frac{1}{4}+V\cdot(\frac{1}{2}-\frac{\sqrt{2}}{4})\nonumber\\
n_6=I_0\cdot\frac{\pi}{8}+\frac{Q}{2}\cdot\frac{\pi}{8}-\frac{Q}{2}\cdot\frac{1}{4}+\frac{U}{2}\cdot\frac{1}{4}+V\cdot\frac{\sqrt{2}}{4}\nonumber\\
n_7=I_0\cdot\frac{\pi}{8}+\frac{Q}{2}\cdot\frac{\pi}{8}-\frac{Q}{2}\cdot\frac{1}{4}-\frac{U}{2}\cdot\frac{1}{4}+V\cdot\frac{\sqrt{2}}{4}\nonumber\\
n_8=I_0\cdot\frac{\pi}{8}+\frac{Q}{2}\cdot\frac{\pi}{8}+\frac{Q}{2}\cdot\frac{1}{4}-\frac{U}{2}\cdot\frac{1}{4}+V\cdot(\frac{1}{2}-\frac{\sqrt{2}}{4})\nonumber
\label{form2}
\end{eqnarray}
Note that the sky background effects should be eliminated from the values $n_1$, $n_2$, $n_3$, $n_4$, $n_5$, $n_6$, $n_7$ and $n_8$. For this, the sky background should be measured prior and after the program object. The mean or interpolated sky background values at the instant of the object observation are subtracted from the observed values for the object for each of the eight channels.

The combination of the pulse counters readings enables to obtain the following dependencies for the Stokes parameters:
\begin{eqnarray}
N=n_1+n_2+n_3+n_4+n_5+n_6+n_7+n_8=\nonumber\\
=I_0\cdot\frac{\pi}{2}\nonumber\\
S_1=-n_1+n_2+n_3+n_4+n_5-n_6-n_7-n_8=\nonumber\\
=-V\cdot\sqrt{2}\nonumber\\
S_2=-n_1-n_2-n_3+n_4+n_5+n_6+n_7-n_8=\\
=V\cdot\sqrt{2}\nonumber\\
S_3=n_1+n_2-n_3-n_4+n_5+n_6-n_7-n_8=U\nonumber\\
S_4=n_1-n_2-n_3+n_4+n_5-n_6-n_7+n_8=Q\nonumber
\label{form3}
\end{eqnarray}
As the parameter $Q\leq0.1$ for the majority of astronomical objects, when neglecting it, the first equation in the set (\ref{form3}) can be written down as $I_0 = N/\pi$. Hence, the standardized parameters $u$ and $q$ of the Stokes vector for the linear polarization expressed in per cent can be determined from the following formulae:
\begin{eqnarray}
LP_1=314.16\cdot\frac{S_3}{N}=\frac{U}{I_0}=u\nonumber\\
LP_2=314.16\cdot\frac{S_4}{N}=\frac{Q}{I_0}=q
\label{form4}
\end{eqnarray}
The Stokes parameters for the circular polarization are determined as follows:
\begin{eqnarray}
CP_1=-S_1\cdot\frac{157.08}{N}\nonumber\\
CP_2=S_2\cdot\frac{157.08}{N}
\label{form5}
\end{eqnarray}
According to the set of equations (\ref{form3}), the sum of pulses accumulated in all channels depends upon the parameter Q. Thus, in general, the second iteration is required to determine the final values of the polarization parameters, but it is only essential when the degree of linear polarization is above 10\%.

The final equations for the degree $p$ and the plane of linear polarization $\theta$ are as follows:
\begin{eqnarray}
p=\frac{\sqrt{u^2+q^2}}{I_0}\nonumber\\
\theta=\frac{1}{2}\arctan\frac{u}{q}
\label{form6}
\end{eqnarray}
The standard error of the circular and linear polarization can be determined from the following formulae:
\begin{eqnarray}
\sigma_v=157.08\frac{\sqrt{N+N_{bgr}}}{N}\nonumber\\
\sigma_p=314.16\frac{\sqrt{N+N_{bgr}}}{N}
\label{form7}
\end{eqnarray}
where $N_{bgr}$ is the total number of the sky background pulses for eight pulse counters.
According to [3], the error in the angle determination can be calculated by the following formula:
$\sigma_{\theta}=28.65\cdot\frac{\sigma_p}{p}$
\\[2mm]

{\bf 2. Determination of the degree of circular polarization by the observed values $CP_1$ and $CP_2$}\\[1mm]

The parameters $CP_1$ and $CP_2$ are equivalent to the Stokes parameters for the linear polarization, namely $q$ and $u$. Therefore, $CP_1$ and $CP_2$ are projections of the vector $P_C$ on the $O_X$ and $O_Y$ axes under the statistical noise perturbations; hence, it is necessary to determine the degree of angle $2\phi_0$.

It can be done using the observations of a star with a wide range of circular polarization variation. To this end, at first, the correction for the zero point from the measured standards with zero polarization should be factored in, i.e. the instrumental polarization should be taken into account. And then, it is necessary to find such an angle of rotation of the coordinate system (i.e. rotation of the polarizer relative to the analyser) that one of the axes corresponds to the polarization while another one represents the noise.\\
Let us introduce a system of coordinates X and Y where the origin coincides with the mean values ($\langle P_1\rangle $, $\langle P_2\rangle $), and the $O_X$ axis is tilted relative to $P_1$ by an angle $\phi$ (which equals to $2\pi\phi_0$). Then,
\begin{eqnarray}
P_1=\overline{P_1}+X\cos\phi-Y\sin\phi\,\,\,\,\,\,\,\,\,\,\,\,\,\,\,\,\,\,\,\,\,\,\,\,\,\,\,\,\,\,\nonumber\\
P_2=\overline{P_2}+X\sin\phi+Y\cos\phi\,\,\,\,\,\,\,\,\,\,\,\,\,\,\,\,\,\,\,\,\,\,\,\,\,\,\,\,\,\,\nonumber\\
X=(P_1-\overline{P_1})\cos\phi+(P_2-\overline{P_2})\sin\phi\,\,\,\,\\
Y=-(P_1-\overline{P_1})\sin\phi+(P_2-\overline{P_2})\cos\phi\nonumber
\label{form8}
\end{eqnarray}
Let us calculate the second central moments for the variables $P_1$ and $P_2$:
\begin{eqnarray}
\mu_{ij}=\langle (P_i-\overline{P_i})^2(P_j-\overline{P_j})^2\rangle \,\,\,\,\,\,\,\,\,\,\,\,\,\,\,\,\,\,\,\,\,\,\,\,\,\,\,\,\,\,\,\,\,\,\,\,\,\,\,\,\,\,\,\,\,\,\,\,\,\,\,\,\,\,\,\,\,\,\,\,\,\,\,\,\,\,\,\,\,\,\nonumber\\
\langle X\rangle =0, \langle Y\rangle =0,\,\,\,\,\,\,\,\,\,\,\,\,\,\,\,\,\,\,\,\,\,\,\,\,\,\,\,\,\,\,\,\,\,\,\,\,\,\,\,\,\,\,\,\,\,\,\,\,\,\,\,\,\,\,\,\,\,\,\,\,\,\,\,\,\,\,\,\,\,\,\,\,\,\,\,\,\,\,\,\,\,\,\,\,\,\,\,\,\,\,\,\,\nonumber\\
\langle X^2\rangle =\mu_{11}\cos^2\phi+2\mu_{12}\cos\phi\sin\phi+\mu_{22}\sin^2\phi=\,\,\,\,\,\,\,\,\,\,\,\,\,\,\nonumber\\
=\frac{1}{2}(\mu_{11}+\mu_{22})+\frac{1}{2}(\mu_{11}-\mu_{22})\cos2\phi+\mu_{12}\sin2\phi\,\,\,\,\,\,\,\,\,\nonumber\\
\langle XY\rangle =-(\mu_{11}-\mu_{22})\cos\phi\sin\phi+
\mu_{12}(\cos^2\phi-\sin^2\phi)=\,\,\,\,\nonumber\\
=-\frac{1}{2}(\mu_{11}-\mu_{22})\sin2\phi+\mu_{12}\cos2\phi\,\,\,\,\,\,\,\,\,\,\,\,\,\,\,\,\,\,\,\,\,\,\,\,\,\,\,\,\,\,\\
\langle Y^2\rangle =\mu_{11}\sin^2\phi+2\mu_{12}\cos\phi\sin\phi+\mu_{22}\cos^2\phi=\,\,\,\,\,\,\,\,\,\,\,\,\,\,\nonumber\\
=\frac{1}{2}(\mu_{11}+\mu_{22})-\frac{1}{2}(\mu_{11}-\mu_{22})\cos2\phi-\mu_{12}\sin2\phi\,\,\,\,\,\,\,\,\,\,\,\,\,\,\nonumber
\label{form9}
\end{eqnarray}
These formulae are valid for any angle of rotation $\phi$, however, for the orthogonal regression, it is required to select such an angle that the joint moment. Thus,
\begin{eqnarray}
\tan2\phi=\frac{2\mu_{12}}{\mu_{11}-\mu_{22}}\,\,\,\,\,\,\,\,\,\,\,\,\,\,\,\,\,\,\,\,\,\,\,\,\,\,\,\,\,\,\,\,\,\,\,\,\,\,\,\,\,\,\,\,\,\,\,\,\,\,\,\,\,\,\,\,\,\,\,\,\,\,\,\,\,\,\,\,\,\,\,\,\,\,\nonumber\\
\sin2\phi=\frac{2\mu_{12}}{\sqrt{(\mu_{11}-\mu_{22})^2+4\mu^2_{12}}}\,\,\,\,\,\,\,\,\,\,\,\,\,\,\,\,\,\,\,\,\,\,\,\,\,\,\,\,\,\,\,\,\,\,\,\,\,\,\,\,\,\,\nonumber\\
\cos2\phi=\frac{\mu_{11}-\mu_{22}}{\sqrt{(\mu_{11}-\mu_{22})^2+4\mu^2_{12}}}\,\,\,\,\,\,\,\,\,\,\,\,\,\,\,\,\,\,\,\,\,\,\,\,\,\,\,\,\,\,\,\,\,\,\,\,\,\,\,\,\,\,\\
\langle X^2\rangle =\frac{1}{2}(\mu_{11}+\mu_{22})+\frac{1}{2}\sqrt{(\mu_{11}-\mu_{22})^2+4\mu^2_{12}}\nonumber\\
\langle XY\rangle =0\,\,\,\,\,\,\,\,\,\,\,\,\,\,\,\,\,\,\,\,\,\,\,\,\,\,\,\,\,\,\,\,\,\,\,\,\,\,\,\,\,\,\,\,\,\,\,\,\,\,\,\,\,\,\,\,\,\,\,\,\,\,\,\,\,\,\,\,\,\,\,\,\,\,\,\,\,\,\,\,\,\,\,\,\,\,\,\,\,\,\,\,\,\,\,\,\,\,\nonumber\\
\langle Y^2\rangle =\frac{1}{2}(\mu_{11}+\mu_{22})-\frac{1}{2}\sqrt{(\mu_{11}-\mu_{22})^2+4\mu^2_{12}}\nonumber
\label{form10}
\end{eqnarray}
In fact, there are four different roots $\phi=\frac{1}{2}\arctan\lambda+\frac{k\pi}{2}$, $k=0,1,2,3$ of the equation $\tan2\phi=\lambda$.

Two of these four roots correspond to the direction of the major axis of the dispersion ellipse along the $O_X$ axis while another two roots correspond to the direction along the $O_Y$ axis. We select the main direction in the quadrant where the formulae for $\sin2\phi$ and $\cos2\phi$ are valid. Then, the variables $X$ and $Y$ have the highest and lowest dispersion, respectively, among all possible angles of rotation. It is rather common that the variable $X$ is interpreted as a variable parameter with the observation noise while $Y$ is interpreted as 'pure noise'.

The total Stokes parameters are calculated in the manner described. It follows from the above-presented formulae that using this technique the circular polarization parameters are determined more precisely than those for the linear polarization.\\[2mm]

{\bf 3. Computer program}\\[1mm]

To process the observation data obtained with the single-channel photopolarimeter, Breus (2007) has written the computer program "PolarObs", which carries out the techniques described above (see also Breus et al., 2007).

A data file is generated by the telescope observations; this file contains eight quasi-simultaneous measurements of the object's brightness for eight successive positions of the modulator and the instant of time at the end of a given observation. Information on the type of the object (such details as dark current, background, target star, reference star and standard), exposure time and number of observation in a series and spectral band-filter is also recorded in the data file by the polarimeter control program, written by a staff member of CrAO, D.N.Shakhovskoy.

The program automatically identifies the type of data sequence using keyword analysis, giving the user the option of accepting or changing the resulted type. Such a fashion allows of minimizing the amount of clicking and key pressing which is more comfortable when processing a large set of data.

\begin{figure}[t]
\label{figure1}
\begin{center}
\resizebox{\hsize}{!}{\includegraphics{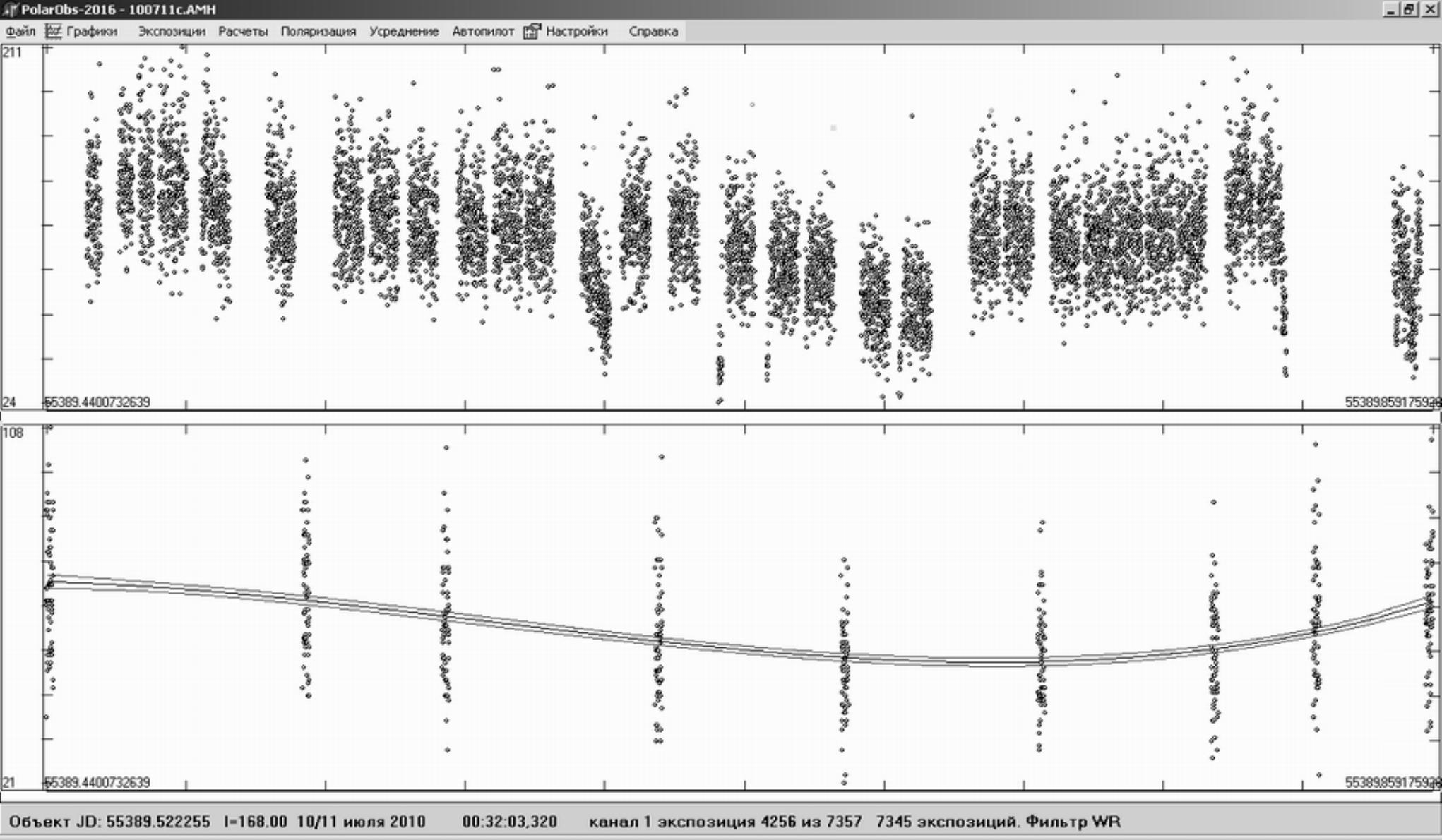}}
\end{center}
\caption{The main screen of Polarobs. Measurements of AM Her at the top and the background at the bottom.}
\end{figure}

When the data set is open, the program exhibits two curves which represent the resulted measurements of the object's brightness (on the top of the screen) and sky background (on the bottom of the screen) for either of eight channels. Having the polynomial fitting of the background counts performed the obtained polynomial values are subtracted from the stars' counts for each channel individually. Subsequently, the user can compute the smoothing polynomial value for the reference star counts in order to determine the brightness of the target object. After that the program computes linear combinations (see \ref{form3}) using fixed constants for each channel which results in the so-called 'vectors' $S_1-S_4$. The first two vectors $S_1$ and $S_2$ are used later to compute the circular polarization parameters while another two vectors ($S_3$ and $S_4$) are used to obtain the linear polarization parameters. Having this step completed, it is possible to save the results obtained in a format of vectors of photometric observation, somehow similar to the Stokes parameters.

As the next step the user can analyse the diagram representing the correlation between $S_2$ and $S_1$ (for the circular polarization) and between $S_4$ and $S_3$ (for the linear polarization). In this view mode the calculated values of polarization, position angle and other data are shown under the diagram.

When processing the standards of zero or non-zero polarization, these data are considered to be and saved as the final results. When processing observations of a variable star or any other object it is necessary to account for the instrumental polarization. To this end, the coordinate system of the linear polarization diagram should be rotated by an angle determined from the standards of non-zero linear polarization.

\begin{figure}[t]
\label{figure2}
\begin{center}
\resizebox{\hsize}{!}{\includegraphics{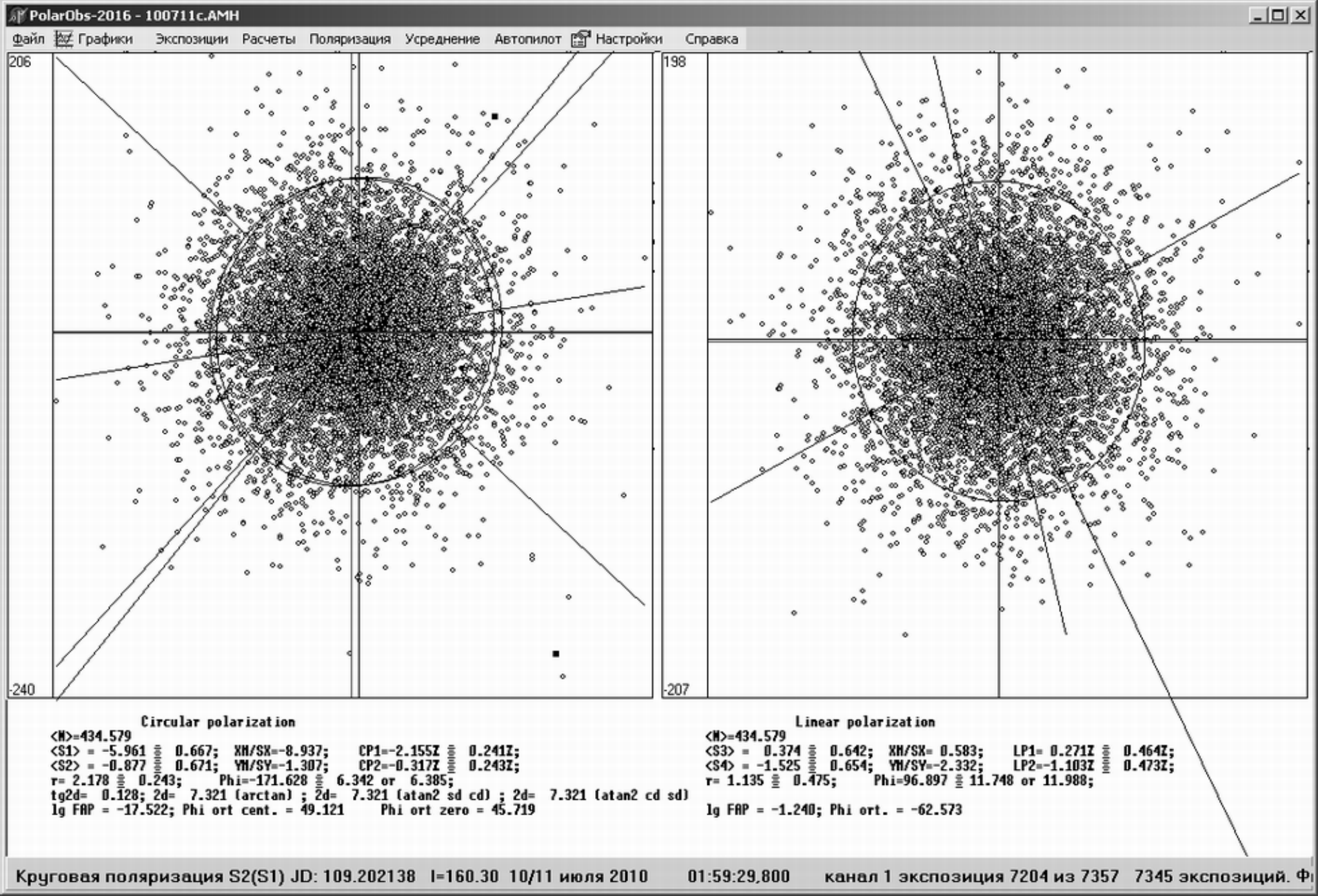}}
\end{center}
\caption{Diagram representing the correlation between $S_2$ and $S_1$ (left) and $S_4$ and $S_3$ (right)}
\end{figure}

To determine the circular polarization, it is necessary to rotate the coordinate system of the circular polarization diagram by a certain angle in such a way that the line connecting the origin and the distribution centre coincides with the $O_X$ axis.

The standardised values of the Stokes parameters can be the output to the files with extension *.p on demand. These are delimited text files with spacebar and newline separated values which contain the following data:

$JD$ is the Julian date;

$F_t$ is the object brightness expressed as a ratio between the object counts and interpolated reference star count;

$F_m$ is the object brightness expressed in magnitude units (related to the Pogson formula Ft);

$CP^\ast_1$, $CP^\ast_2$ are the circular polarization values;

$LP^\ast_1$, $LP^\ast_2$ are the linear polarization values;

$\sigma_{F}$, $\sigma_{CP}$, $\sigma_{LP}$ are the errors in photometry, circular and linear polarization, respectively.

In the last step it is possible to perform either polynomial approximation or averaging of the standardised or not-standardised Stokes parameters. When averaging, the program gives an option to select statistically optimal number of points for averaging using three test-functions, such as the estimated error of a single measurement, average accuracy of the smoothed value and the signal-to-noise ratio.

\begin{figure}[h]
\label{figure3}
\begin{center}
\resizebox{\hsize}{!}{\includegraphics{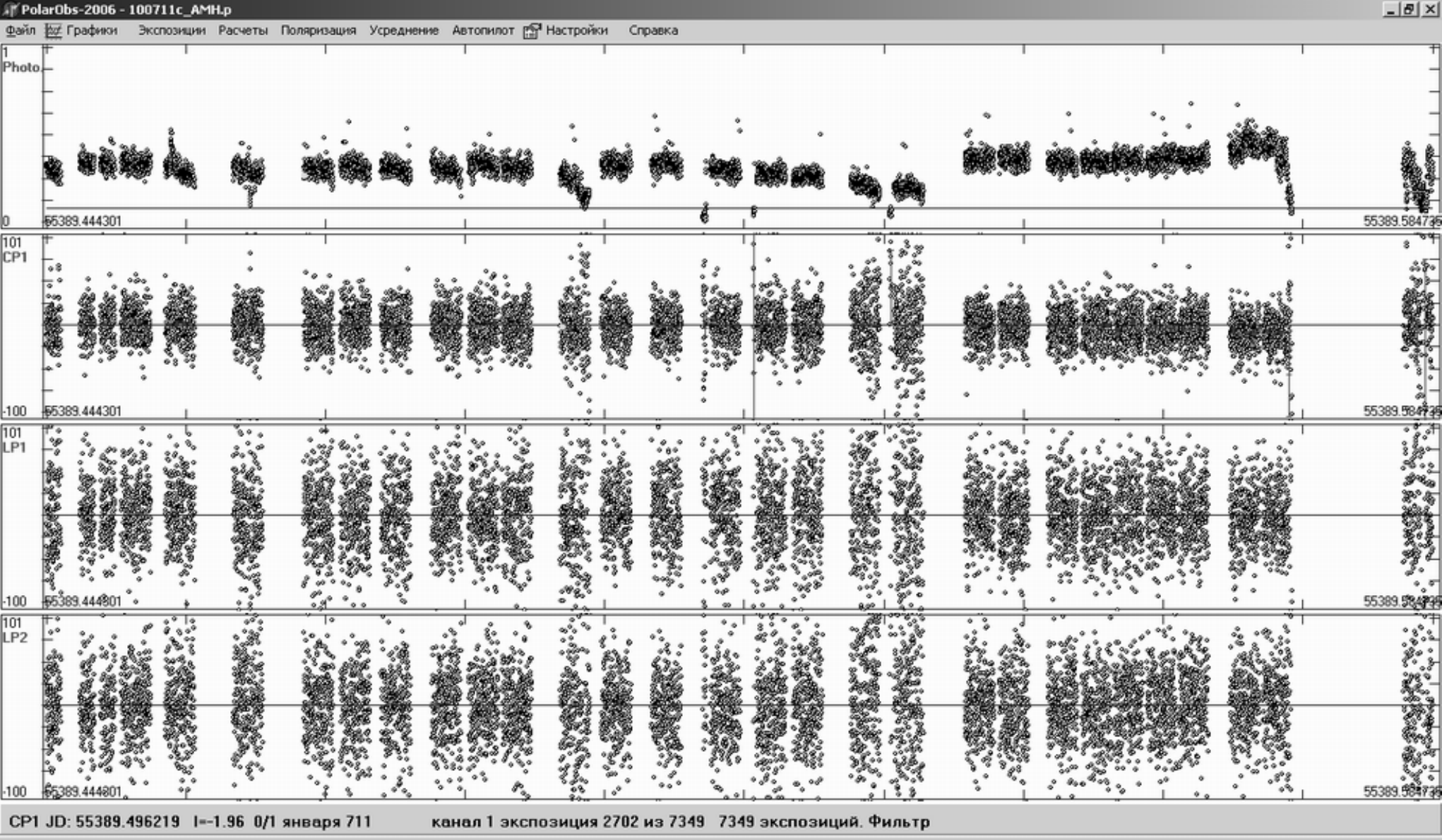}}
\end{center}
\caption{Viewing P-file - photometry (1), circular (2) and linear (3-4) polarization}
\end{figure}

The program was used to process the photopolarimetric observations of the stellar systems V405 Aur (Breus et al., 2013), BY Cam (Andronov et al., 2008), AM Her (Andronov et al., 2003) and QQ Vul (Andronov et al., 2010), as well as several comets (Kiselev et al., 2012; Rozenbush et al., 2007, 2009, 2014). Some results were reported in reviews on large scientific campaigns (Andronov et al., 2010; Vavilova et al., 2011, 2012).\\[2mm]

\newpage\indent
{\bf References\\[2mm]}
Andronov I.L. et al.: 2003, {\it Odessa\,Astron.\,Publ.,}\,{\bf 16},\,7.\h 2003OAP....16....7A\\
Andronov I.L. et al.: 2008, {\it Central European Journal\h of Physics,} {\bf 6(3)}, 385. 2008CEJPh...6..385A\\
Andronov I.L. et al.: 2010, {\it Odessa\,Astron.\,Publ.,}\,{\bf 23},\,8.\h 2010OAP....23....8A\\
Breus V.V. : 2007, {\it Odessa Astron. Publ.,} {\bf 20}, 32.\h 2007OAP....20...32B\\
Breus V.V., Andronov I.L., Kolesnikov S.V., Shakho-\h vskoy N.M.: 2007, {\it AATr}, {\bf 26}, 241.\h 2007A\&AT...26..241B\\
Breus V. V. et al.: 2013, {\it JPhSt}, {\bf 17}, 3901.\h 2013JPhSt..17.3901B \\
Kiselev N.N. et al.: 2012, {\it LPI Contr.} {\bf 1667}, 6102.\h 2012LPICo1667.6102K\\
Kucherov V.A.: 1986, {\it KPCB}, {\bf 2}, 59.\h 1986KFNT....2...59K\\
Rozenbush V.K. et al.: 2007, {\it The 10-th Conference\h on Electromagnetic and Light Scattering, Bodrum,\h Turkey, Ed. by G.Videen et al.} 2007Icar..186..317R\\
Rozenbush V.K. et al.: 2009, {\it Journal of\h Quantitative Spectroscopy and Radiative Transfer},\h {\bf 110 (14)}, p. 1719. 2009JQSRT.110.1719R\\
Rosenbush V. et al.: 2014, {\it Asteroids, Comets,\h Meteors, Helsinki, Finland. Edited by\h K. Muinonen et al.}, p. 450. 2014acm..conf..450R\\
Serkowski K.: 1974. {\it Planets, Stars and Nebulae,\h Studied with Photopolarimetry, ed. by Gehrels,\h Univ. of Arizona Press, Tucson}, p. 135.\h 1974psns.coll..135S\\
Shakhovskoy N.M. et al.: 2001, {\it IzKry}, {\bf 97}, 91.\h 2001IzKry..97...91S \\
Vavilova I.B. et al.: 2011, {\it KosNT}, {\bf 17}, 74.\h 2011KosNT..17d..74V\\
Vavilova I.B. et al.: 2012, {\it KPCB}, {\bf 28}, 85.\h 2012KPCB...28...85V\\

\vfill
%

\end{document}